\documentclass[conference,a4paper]{IEEEtran}
\IEEEoverridecommandlockouts
\usepackage{cite}
\usepackage{amsmath}
\usepackage{mathtools}
\usepackage{booktabs}
\usepackage{graphicx}
\usepackage{amsmath,amssymb,amsfonts}
\usepackage{algorithm, algpseudocode}
\usepackage{graphicx}
\usepackage{float}
\usepackage{textcomp}
\usepackage[table,xcdraw]{xcolor}
\usepackage{hyperref}
\hypersetup{draft}
\def\BibTeX{{\rm B\kern-.05em{\sc \kern-.025em b}\kern-.08em
    T\kern-.1667em\lower.7ex\hbox{E}\kern-.125emX}}

\begin{document}

\title{Utilizing Collaborative Filtering in a Personalized Research-Paper Recommendation System}

\author{

\IEEEauthorblockN{Mahamudul Hasan}
\IEEEauthorblockA{\textit{Computer Science and Engineering} \\
\textit{University of Minnesota Twin Cities}\\
Minneapolis, United States\\
munna09bd@gmail.com}

\and

\IEEEauthorblockN{Anika Tasnim Islam}
\IEEEauthorblockA{\textit{Computer Science and Engineering} \\
\textit{East West University}\\
Dhaka, Bangladesh \\
anikashukh@gmail.com}

\and
\IEEEauthorblockN{Nabila Islam}
\IEEEauthorblockA{\textit{Computer Science and Engineering} \\
\textit{East West University}\\
Dhaka, Bangladesh \\
mail2nabilaislam@gmail.com}

}

\maketitle

\begin{abstract}
Recommendation system is such a platform that helps people to easily find out the things they need within a few seconds. It is implemented based on the preferences of similar users or items. In this digital era, the internet has provided us with huge opportunities to use a lot of open resources for our own needs. But there are too many resources on the internet from which finding the precise one is a difficult job. Recommendation system has made this easier for people. Research-paper recommendation system is a system that is developed for people with common research interests using a collaborative filtering recommender system. In this paper, coauthor, keyword, reference, and common citation similarities are calculated using Jaccard Similarity to find the final similarity and to find the top-n similar users. Based on the test of top-n similar users of the target user research paper recommendations have been made. Finally, the accuracy of our recommendation system has been calculated. An impressive result has been found using our proposed system.
\end{abstract}


\section{Introduction}
 
From its inception, the field of recommendation systems has progressed through both essential research and business advancement to where today recommender frameworks are implanted in a wide range of content applications. 


Recommendation system provides suggestions about the item list to the user. It refers to a system that predicts future preferences of an item set for the user. Because these days people have so many options from which to choose. Recommendation systems have made this decision-making easier. Previously people used to go to stores where the availability of items is limited. By contrast, recently people can access abundant resources online. But, a new problem arises as they have hardly too many times to choose from so many options. The concept of a recommendation system has come from this scenario.\\ 
There exist a lot of different recommendation system approaches with their limitations. The first one is collaborative filtering (CF) which is considered to be the most successful recommendation approach as it refers to the recommendation based on ratings given by similar users. The motive of collaborative filtering is to introduce new items to a user based on the recommendation of others' who shares a similar interest. Collaborative Filtering is of two types: memory-based collaborative filtering and model-based collaborative filtering \cite{adomavicius2005toward}.
An example of memory based collaborative filtering is the User-Based collaborative filtering algorithm where it considers all the previous rated items of a user to predict. Model-based collaborative filtering considers a model from a set of ratings and uses this to predict. An example of model-based collaborative filtering is item-based collaborative filtering. This type of approach is more scalable than memory-based CF. Another recommendation approach utilizes demographic attributes of user with the help of pre-generated demographic clusters to recommend. Users obtain recommendation systems on the basis of ratings or search histories. But when a new user comes to the system, his preferences history remains empty, in that case, it is difficult for the system to recommend. This problem can be stated as a cold-start problem \cite{maltz1995pointing}.\\
Content-based filtering is another approach that predicts based on user preferences.
By using the similarity of the user-searched keyword with other keywords and the user preferences in some cases recommendation can be made \cite{liu2015context}.\\
This paper mainly focuses on collaborative filtering recommendation systems. The core part of collaborative filtering is the calculation of similarities among users and then predicts the appropriate items based on the top-n similarities. Also, it builds a model from the user's past behavior preferences to the item sets. It has come up with such a recommendation system that recommends the most suitable paper to the researcher by calculating paper similarity on user preference.

\section{Literature Review}


Collaborative filtering is recommended based on the past preferences of the user and it is believed that the past user preference will remain the same in the present and future, they will prefer items which is similar to the preferred items in the past.
Though it has been widely used, it is prone to many problems like cold-start problems, sparsity of data, scalability, and many others. Many researchers have proposed various solutions to enhance the precision value. Such as Ahn et al \cite{liu2014new} proposed PIP(Proximity-Impact-Popularity), a new similarity measure for a recommender system using a collaborative filtering approach. Bobadilla et al \cite{bobadilla2012collaborative} paired the Jaccard measure and mean square difference. Also, MJD(Mean-Jaccard-Difference) was created with the intention of solving the cold-start problem. 
Another method for improving the recommender system in collaborative filtering is data smoothing. Also, some other techniques such as neural networks, and zero-sum reward support vector machines are implemented for the improvement of accuracy but these are not efficient enough for solving the cold-start problem.



In \cite{liu2015context} they have implemented a model using citation context-based collaborating filtering. The model of this paper has been designed in terms of citing papers, cited papers, and co-occurrences of citing papers and created a binary rating matrix and paper vectors which represented association between citing papers.

In \cite{hong2012userprofile} A model has been implemented by extracting keywords from papers to generate a user profile and also updating the user profile continuously by extracting the keyword. Natural language processing techniques have been used to get appropriate representation words from the paper that doesn't have any keywords. 

In \cite{hong2013personalized} they have implemented a User Profile-based algorithm by keyword interference and extraction of keyword. If any paper doesn't have any keyword section, this system considers the title as a keyword and execute the algorithm.

In \cite{zhang2010research} a personalized research paper recommendation system has been proposed that constructs a user profile based on a concept tree in order to overcome the flaws of the classic vector space model. They have used the spreading activation model to search for users sharing similar interests.

In \cite{haruna2017collaborative} an evaluation system for research-paper recommender system based on collaborative filtering has been proposed. Here, they have used collaborative filtering to form their model and mined the hidden association between research papers. The work was based on all of the references and citations of a target paper. In the proposed approach, they built a paper citation relation matrix between reference and citation to generate the recommendation systems.

\section{Proposed Section}
\begin{figure}[H]
  \centering
  \includegraphics[width=\linewidth]{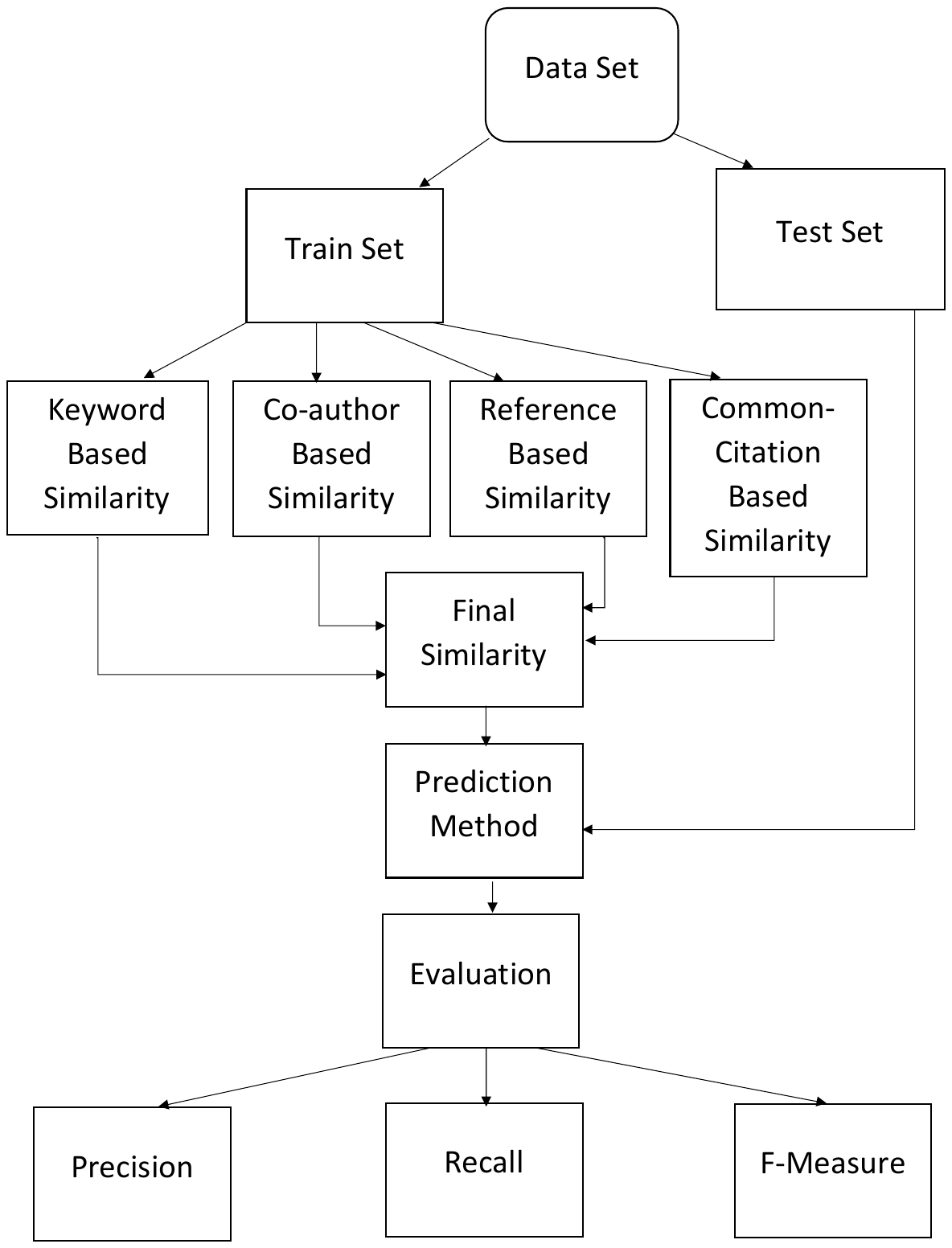}
  \caption{Flowchart of our proposed model}
  \label{gg}
\end{figure}

A user-based collaborative filtering has been applied in our proposed system. Every user in the system has some publication and the final recommendation has been made based on the preferences of the users. A user will be recommended relevant papers according to his preferred keywords, co-authors, references, and citations. The system will find similar users based on the common keywords, co-authors, references, and citations. The research paper recommendations will be made based on a similar user's reference list. Ultimately, the accuracy of our system has been evaluated. The experimental section shows that the performance of our system is excellent in terms of precision, recall, and f-measures. The flow chart of our system has been shown in the figure \ref{gg}. 

The dataset has been split into the testing and training sets. Then the keyword-based similarity, co-author-based similarity, reference-based similarity, and common citation-based similarity have been calculated. After that, we have designed a unified similarity by which we can find the final similarity. Then using the test data the prediction has been done using the final similarity. Finally, the accuracy of the system has been calculated.
Our proposed approach improves the performance of the overall scholarly paper recommendation system. The procedures are elaborated in \ref{AA}.
\subsection{Calculating User-User Similarity}\label{AA}
In this section, the similarity of the users is considered. For this system, to calculate the similarity between two users, Jaccard similarity to find similarities between keyword, co-author, common-citation, and reference have been used. Then we put those values into the similarity function, assigned weights with each similarity and then the final similarity matrix has been found. From that final similarity matrix, we chose top-n similar users. The process is described elaborately in the sub-section a. Similarity between keywords, b. Similarity between co-authors, c. The similarity between citations, and d. Similarity between common references.  

\paragraph{Calculating Keyword Similarity}\label{CC}
For our proposed method, first, we implement keyword-based similarity using Jaccard similarity where our proposed equation is \ref{equ1}.

\begin{equation}
Sim_{Key}(U,V)=\frac{\mid U_{keyword}\cap V_{keyword} \mid}{\mid U_{keyword}\cup V_{keyword}\mid} \label{equ1}\\
\end{equation}

It calculates the keyword-based similarity between two users U and V. We extract all the keywords from an author's papers and compare them to find this keyword-based similarity.

\paragraph{Calculating Co-author Similarity}\label{DD}
Now, we calculate the similarity between two co-authors using the Jaccard Similarity where our proposed equation is \ref{equ2}.

\begin{equation}
Sim_{Co-auth}(U,V)=\frac{\mid U_{Co-author}\cap V_{Co-author}\mid}{\mid
U_{Co-author}\cup V_{Co-author}\mid} \label{equ2}\\
\end{equation}

Users who contribute to the same paper are considered co-authors. When the number of papers increases as co-authors between two users then we can conclude that both users are similar.

\paragraph{Calculating Common-citation Similarity}\label{EE}
After that, we calculate how frequently two users have got common citations. This similarity is also measured using Jaccard Similarity where our proposed equation is \ref{equ3}.

\begin{equation}
Sim_{cit}(U,V)=\frac{\mid U_{Citation}\cap V_{Citation}\mid}{\mid U_{Citation}\cup V_{Citation}\mid} \label{equ3}\\
\end{equation}

Users who often cite the papers of each other are more likely to be working in the same research field. This way we can also find similarities between two users.

\paragraph{Calculating Reference Similarity}\label{FF}
And at last, we calculate the reference similarity between two users. Our proposed equation for calculating this similarity is \ref{equ4}.

\begin{equation}
Sim_{ref}(U,V)=\frac{\mid U_{reference}\cap V_{reference}\mid}{\mid U_{reference}\cup V_{reference}\mid} \label{equ4}\\
\end{equation}

Reference similarity plays an important role here. By this similarity, we can find the frequent reference behavior between two users.

\subsection{Similarity Function}
An ensemble similarity has been proposed based on the individual similarity with a weighted factor that are depicted in figure \ref{equ5}.

\begin{equation}
Sim(U,V)=\left[{\dfrac{\splitdfrac{\alpha \times Sim_{key}+\beta \times Sim_{ref}}{+\gamma \times Sim_{Co-auth}+\mu \times Sim_{cit}}}{\alpha + \beta + \gamma + \mu}} 
\right]
\label{equ5} 
\end{equation}

where, 
$\alpha+\beta +\gamma +\mu = 1$

\subsection{Similarity Matrix}

From the score generated from the similarity function, a sample similarity matrix will be formed and our similarity matrix has been shown in the table. 

\begin{table}[h]
\caption{Similarity Matrix Table}
\begin{tabular}{|>{\columncolor[HTML]{C0C0C0}}l |l|l|l|l|l|}
\toprule
User/User                 & \cellcolor[HTML]{C0C0C0}U1 & \cellcolor[HTML]{C0C0C0}U2 & \cellcolor[HTML]{C0C0C0}U3 & \cellcolor[HTML]{C0C0C0}U4 & \cellcolor[HTML]{C0C0C0}U5  \\ \midrule
U1                        & -                          & 0.09                        & 0.03                        & 0.03                        & 0.04                         \\ \midrule
U2                        & -                          & -                          & 0.02                        & 0.08                        & 0.09                         \\ \midrule
U3                        & -                          & -                          & -                          & 0.07                        & 0.09                         \\ \midrule
U4               & -                          & -                          & -                          & -                          & 0.08 \\ \midrule
U5                        & -                          & -                          & -                          & -                          & - \\\bottomrule

\end{tabular}

\end{table}

From the majority of scores, top-n similar users are taken for the recommendation of the research paper.

\subsection{Paper Recommendation}
In this subsection the paper recommendation algorithm \ref{alg:loop} has been discussed. For any user, we have to find the list of papers that can be recommended. For each user based on his similarity, we have to find the top-n similar users. When any paper has been preferred by majority of the similar users then that paper can be a candidate for the recommendation. Randomly a paper list has been selected then for each paper on the list, when a similar user cited that paper then the counter will be incremented. The paper will become the candidate of the recommended list when the counter reaches a minimum threshold limit. 
Finally, the order of the list has been calculated based on the counter value in descending order.
\renewcommand{\algorithmicrequire}{\textbf{Input:}}
\renewcommand{\algorithmicensure}{\textbf{Output:}}
\begin{algorithm}[h] 
\caption{Paper recommendation algorithm}
\label{alg:loop}
\begin{algorithmic}[1]
\Require{$Users ~and ~paper ~list$} 
\Ensure{$Recommended ~Papers$}
\Statex
\For{$paper \in Paper List$} 
    \For{$u \in U$} 
        \State {$count \gets 0$}
        \For{$v \in U$} 
            \If{$ v \in TopN \left[ sim(u,v)\right]$}
            
                \If{$Ref[v]==paper$}
                    \State {$count \gets count + 1$} 
                \EndIf
            \EndIf
         \EndFor
        \State {$Majority \gets \frac{(count \times 100 )}{TopN}$}
        \If{$\left(Majority \ge minimum\: threshold \right)$} 
            \State {$Recommended\: list \gets Paper$} 
        \EndIf
    \EndFor
\EndFor
\end{algorithmic}
\end{algorithm}

\section{Empirical Analysis}

\subsection{Data-set}
For our experiment, data set has been taken for this \cite{semantic} website. This data set contains a collection of more than 20GB of data. We have to pre-process the data so that the required attributes for this research can be taken from the data set. We have taken the paper-id, user-id, keyword, reference, citations, and co-author information. ten-fold cross-validation has been used while measuring the accuracy of the system.

\subsection{The Evaluation Metrics}
We have used the three most popular evaluation metrics. These are precision, recall, and f-measure. These metrics assist users in identifying papers of very high preference from the range of available research papers \cite{sarwar1998using}.
\\

\paragraph{Precision}
Precision is a measure of accuracy that assesses the percentage of relevant items from all items retrieved. True positive is another term that is defined as the papers which are predicted in the recommended list and they are also actually in the reference list of the target users. The approach for calculating the precision is given in equation \ref{equ7}.

\begin{equation}
    Precision = \frac{True\:Positive}{True\:Positive+False\:Positive}
    \label{equ7}
\end{equation}
\\
\paragraph{Recall}

Recall is a completeness metric that determines the percentage of the retrieved relevant items collected from all the relevant items. False negative is a term that is evaluated as true when the relevant papers are predicted negatively. The formula of recall is given in equation  \ref{equ8}.
\begin{equation}
    Recall= \frac{True\:Positive}{True\:Positive+False\:Negative}
    \label{equ8}
\end{equation}

\paragraph{F-Measure}

The f-measure metric merges precision and recall into a single value in order to compare. It is used for a more balanced view of performance. The formula of F-measure is given in equation \ref{equ9}.

\begin{equation}
    F_{Measures}=\frac{2\times Precision \times Recall}{Precision + Recall}
    \label{equ9}
\end{equation}

\section{Result}

For experimental purposes, we have taken top-n neighbors and compared them with Precision, Recall, and F-measure which helped us to compare that our proposed approach is more efficient than the other approaches proposed or being implemented previously. Collaborative filtering is a renowned and most commonly used approach for recommendation systems. This approach made our model more scalable because of item-based filtering. However, the improvement is not significantly large. Although the system ensures high-performance \cite{sarwar2001item}.

\begin{figure}[H]
  \centering
  \includegraphics[width=\linewidth]{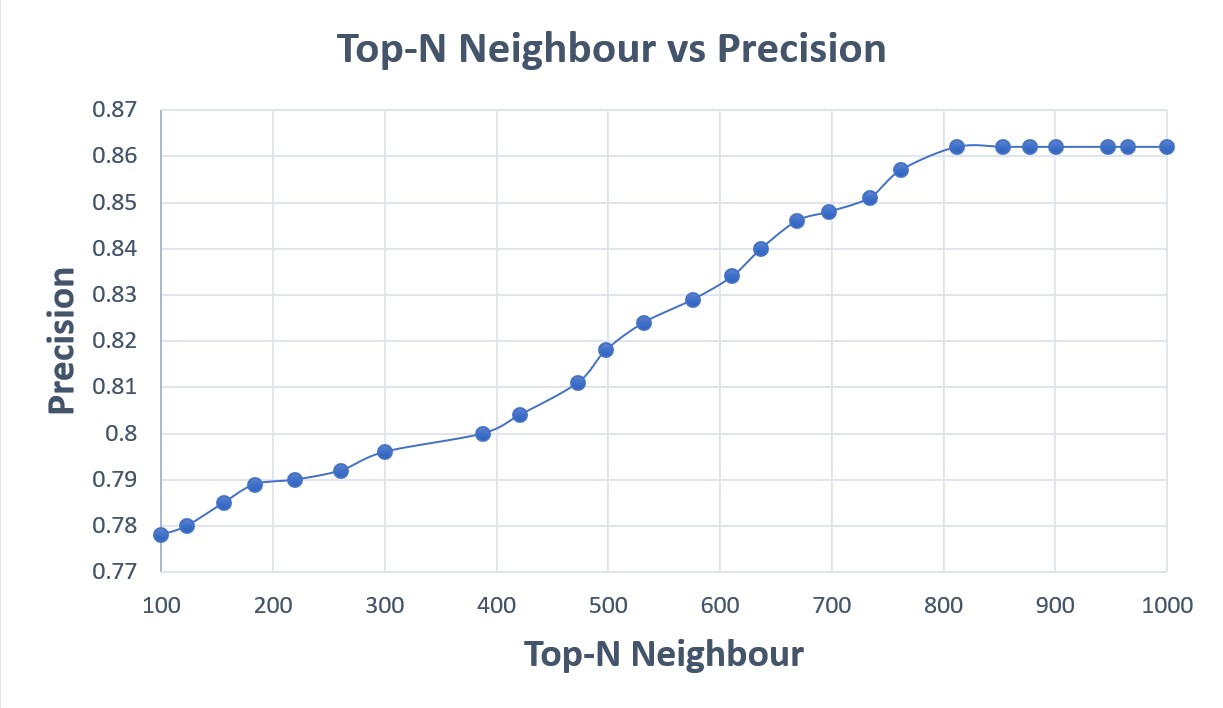}
  \caption{Top-N Neighbours Vs. Precision}
\end{figure}

Precision can be thought of as a degree of exactness. That means the selection of data points classified as positive are positive in reality. If a precision value is 1 that means every data point that is classified positive, is positive indeed. Here, in the graph, we can see that for 100 top-N neighbors, the precision value is 0.779, and for 200 it is 0.79 which means it keeps increasing if the top-N neighbor number gets higher. As it comes to top-1000 neighbors the fraction becomes 0.862 which means the larger the top-N neighbor the greater the precision fraction. This signifies that the proposed approach gives a result that is close to the exact one. 

\begin{figure}[H]
  \centering
  \includegraphics[width=\linewidth]{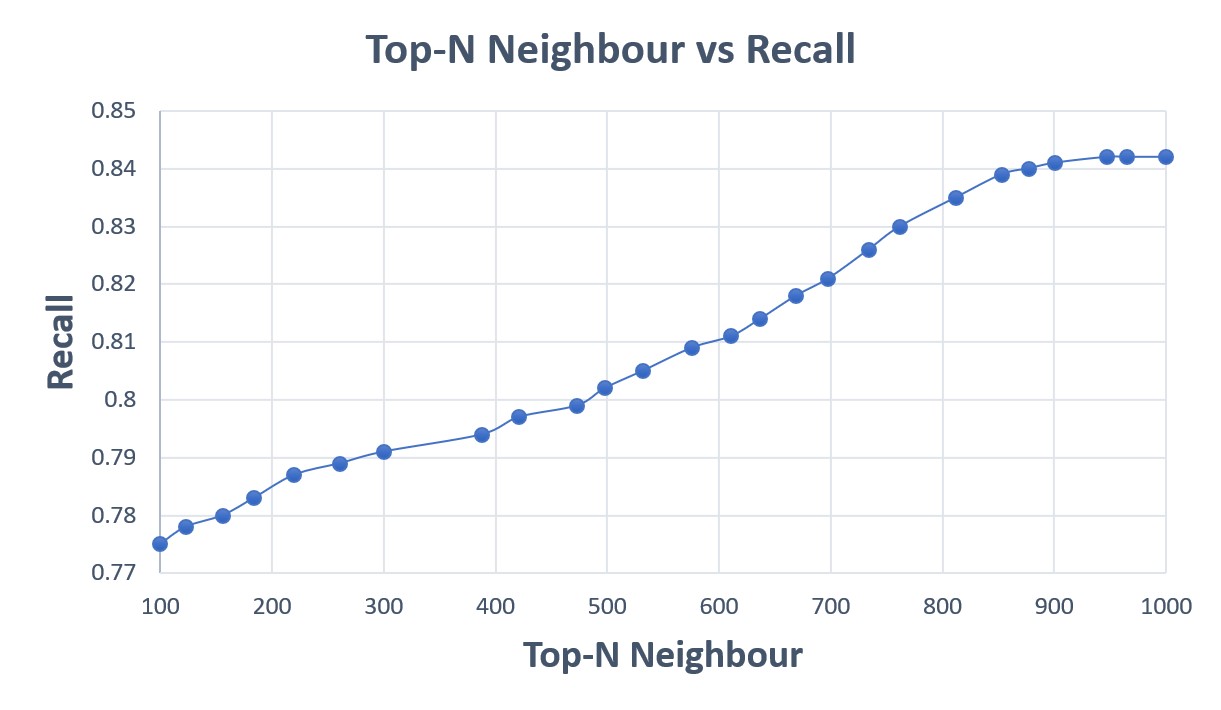}
  \caption{Top-N Neighbours Vs. Recall}
\end{figure}

Recall is a degree of completeness which means positive data points are classified as positive. If a recall value is 1 that means all of the data points which were actually positive are classified as positive. In the graph here, it is clear that for the top-100 neighbor the recall value is 0.777 and for the top-200 neighbor the value becomes 0.785 which means when the top-N neighbor increases, the recall value also gets higher. As the value of top-N neighbor reaches to 1000 the value of recall becomes 0.841 which means if our data had more number of values then we could have reached the highest recall value. This signifies our proposed approach to be more effective than all other approaches that were previously proposed.

\begin{figure}[H]
  \centering
  \includegraphics[width=\linewidth]{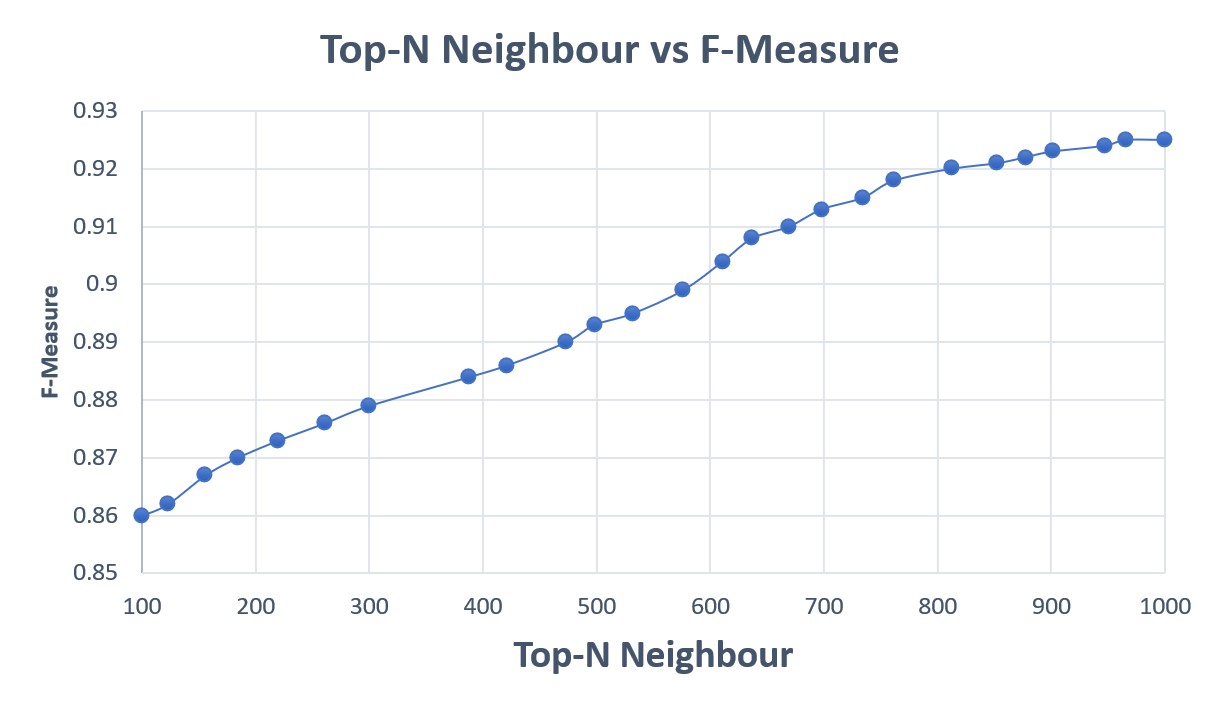}
  \caption{Top-N Neighbours Vs. F-measure}
\end{figure}

The F-measure is known to be the weighted harmonic mean of the value of recall and precision. Without the impact of true negatives that made the accuracy aimless for assessing search algorithms, the f-measure can be thought of as a degree of accuracy. It passes on the balance between precision and recall \cite{haruna2017collaborative}. The lowest possible value of the f-measure is zero, if either of precision or recall value becomes zero. F-measure can moreover be deciphered as the dice coefficient between the retrieved and the relevant set. It is also known as the dice-similarity coefficient. For top 1000 neighbors in the graph, the value is higher than the value of the top 100 neighbors which means the more the user number is, the more balanced the proposed system will be.

\section{Conclusion}
In this paper, collaborative filtering with different measures has been used. Through experiments, we tried to prove the effectiveness of our proposed approach. The limitation of this proposed approach is we didn't consider the public contextual contents for extracting the recommended paper, like the title, and the abstract. In the future, the similarity between abstract and title can be used to get more accurate recommendations. A deep learning-based approach like the recurrent neural network is another solution for text-based similarity and to get accurate recommendations. 
\bibliography{bibliography}

\begin{thebibliography}{10}
\providecommand{\url}[1]{#1}
\csname url@samestyle\endcsname
\providecommand{\newblock}{\relax}
\providecommand{\bibinfo}[2]{#2}
\providecommand{\BIBentrySTDinterwordspacing}{\spaceskip=0pt\relax}
\providecommand{\BIBentryALTinterwordstretchfactor}{4}
\providecommand{\BIBentryALTinterwordspacing}{\spaceskip=\fontdimen2\font plus
\BIBentryALTinterwordstretchfactor\fontdimen3\font minus \fontdimen4\font\relax}
\providecommand{\BIBforeignlanguage}[2]{{%
\expandafter\ifx\csname l@#1\endcsname\relax
\typeout{** WARNING: IEEEtran.bst: No hyphenation pattern has been}%
\typeout{** loaded for the language `#1'. Using the pattern for}%
\typeout{** the default language instead.}%
\else
\language=\csname l@#1\endcsname
\fi
#2}}
\providecommand{\BIBdecl}{\relax}
\BIBdecl

\bibitem{adomavicius2005toward}
G.~Adomavicius and A.~Tuzhilin, ``Toward the next generation of recommender systems: A survey of the state-of-the-art and possible extensions,'' \emph{IEEE transactions on knowledge and data engineering}, vol.~17, no.~6, pp. 734--749, 2005.

\bibitem{maltz1995pointing}
D.~Maltz and K.~Ehrlich, ``Pointing the way: active collaborative filtering,'' in \emph{Proceedings of the SIGCHI conference on Human factors in computing systems}, 1995, pp. 202--209.

\bibitem{liu2015context}
H.~Liu, X.~Kong, X.~Bai, W.~Wang, T.~M. Bekele, and F.~Xia, ``Context-based collaborative filtering for citation recommendation,'' \emph{IEEE Access}, vol.~3, pp. 1695--1703, 2015.

\bibitem{liu2014new}
H.~Liu, Z.~Hu, A.~Mian, H.~Tian, and X.~Zhu, ``A new user similarity model to improve the accuracy of collaborative filtering,'' \emph{Knowledge-Based Systems}, vol.~56, pp. 156--166, 2014.

\bibitem{bobadilla2012collaborative}
J.~Bobadilla, F.~Ortega, A.~Hernando, and J.~Bernal, ``A collaborative filtering approach to mitigate the new user cold start problem,'' \emph{Knowledge-Based Systems}, vol.~26, pp. 225--238, 2012.

\bibitem{hong2012userprofile}
K.~Hong, H.~Jeon, and C.~Jeon, ``User profile-based personalized research paper recommendation system,'' in \emph{2012 8th International Conference on Computing and Networking Technology (INC, ICCIS and ICMIC)}.\hskip 1em plus 0.5em minus 0.4em\relax IEEE, 2012, pp. 134--138.

\bibitem{hong2013personalized}
------, ``Personalized research paper recommendation system using keyword extraction based on user profile,'' \emph{Journal of Convergence Information Technology}, vol.~8, no.~16, p. 106, 2013.

\bibitem{zhang2010research}
Z.~Zhang and L.~Li, ``A research paper recommender system based on spreading activation model,'' in \emph{The 2nd International Conference on Information Science and Engineering}.\hskip 1em plus 0.5em minus 0.4em\relax IEEE, 2010, pp. 928--931.

\bibitem{haruna2017collaborative}
K.~Haruna, M.~A. Ismail, D.~Damiasih, J.~Sutopo, and T.~Herawan, ``A collaborative approach for research paper recommender system,'' \emph{PloS one}, vol.~12, no.~10, 2017.

\bibitem{semantic}
``http://s2-public-api-prod.us-west-2.elasticbeanstalk.com/corpus/download/.''

\bibitem{sarwar1998using}
B.~M. Sarwar, J.~A. Konstan, A.~Borchers, J.~Herlocker, B.~Miller, and J.~Riedl, ``Using filtering agents to improve prediction quality in the grouplens research collaborative filtering system,'' in \emph{in the GroupLens Research Collaborative Filtering System???. Proceedings of the ACM Conference on Computer Supported Cooperative Work (CSCW}, 1998.

\bibitem{sarwar2001item}
B.~M. Sarwar, G.~Karypis, J.~A. Konstan, J.~Riedl \emph{et~al.}, ``Item-based collaborative filtering recommendation algorithms.'' \emph{Www}, vol.~1, pp. 285--295, 2001.

\end{thebibliography}
\bibliographystyle{IEEEtran}

\end{document}